\documentstyle[floats,aps,epsf,amsfonts]{revtex}
\def\lesssim{\mathrel{\hbox{\rlap{\hbox{\lower4pt\hbox{$\sim$}}}\hbox{$<$}}}}
\def\gtrsim{\mathrel{\hbox{\rlap{\hbox{\lower4pt\hbox{$\sim$}}}\hbox{$>$}}}}
\newcommand {\be}       {\begin{equation}}
\newcommand {\ee}       {\end{equation}}
\tighten
\begin{document}
\draft

\title{Confusion Noise from Extreme-Mass-Ratio Inspirals}

%*******************************************************************
\author{Leor Barack$^1$ and Curt Cutler$^2$}
\address
%{$^1$Department of Physics and Astronomy and
%Center for Gravitational Wave Astronomy,
%University of Texas at Brownsville,
%Brownsville, Texas 78520\\
{$^1$Department of Mathematics, University of Southampton, Southampton,
SO17 1BJ, United Kingdom\\
$^2$Max-Planck-Institut f\"{u}r Gravitationsphysik,
Albert-Einstein-Institut, Am M\"{u}hlenberg 1,
D-14476 Golm bei Potsdam, Germany}
%*******************************************************************

\date{\today}
\maketitle

\begin{abstract}

Captures of compact objects (COs) by massive black holes in galactic
nuclei (aka ``extreme-mass-ratio inspirals'') will be an important source for LISA.
However, a large fraction of captures will not be individually resolvable
%---either
%because they are too distant, have unfavorable orientation, or have too many years
%to go before final plunge---
and so will constitute a source of ``confusion noise,''
obscuring other types of sources.  Here we estimate the shape and overall magnitude
of the spectrum of confusion noise from CO captures. The overall magnitude depends
on the capture rates, which are rather uncertain, so we present results for a plausible
range of rates. We show that the impact of capture confusion noise on the total LISA
noise curve ranges from insignificant to modest, depending on these rates.
Capture rates at the high end of estimated ranges would raise LISA's overall
(effective) noise level by at most a factor $\sim 2$.
While this would somewhat decrease LISA's sensitivity to other
classes of sources, it would be a pleasant problem for LISA to have, overall,
as it would also imply that detection rates for CO captures were at nearly their
maximum possible levels (given LISA's baseline design).

\end{abstract}

\pacs{04.80.Nn, 04.30.Db}

Over its projected 3--5 yrs mission time, LISA will detect hundreds to
thousands of stellar-mass compact objects (COs) captured by massive
black holes (MBHs) in galactic nuclei (events aka ``extreme-mass ratio
inspirals''), with detection rate dominated by captures of
$\sim 10 M_{\odot}$ black holes (BHs) \cite{emri}. Detection of individual captures
will be challenging computationally (except for the very close ones \cite{LQ}),
due to the huge parameter
space of capture waveforms. A first-cut data analysis scheme assuming realistic
computational resources was recently presented in \cite{emri}. In this contribution
we point out that most captures of white dwarfs (WDs) and neutron stars (NSs),
as well as a fair fraction of BH captures, will not be individually resolvable,
and hence will constitute a source of {\it confusion noise}, obscuring other
types of sources. We estimate the shape and overall magnitude of the GW energy
spectrum from this capture background, from which we derive an effective spectral
density ${\cal S}^{\rm capt}_h(f)$ for the amplitude of capture-generated
gravitational waves (GWs) registered
by LISA. We then estimate what fraction of ${\cal S}^{\rm capt}_h(f)$
comes from unresolvable sources, and so represents confusion noise.
Our full analysis is provided in Ref.\ \cite{BCconf}. Here we
review the basic steps of this analysis, and summarize the main results.
Unless otherwise indicated, we use units in which $G=c=1$.

Captures occur when two objects in the dense stellar cusp surrounding a galactic
MBH undergo a close encounter, sending one of them into an orbit tight enough
that orbital decay through emission of GWs dominates the
subsequent evolution. For a typical capture, the initial orbital eccentricity
is extremely large (typically $1-e\sim 10^{-6}{-}10^{-3}$) and the initial
pericenter distance very small ($r_{\rm p}\sim 8-100M$, where $M$ is the
MBH mass)~\cite{FreitagApJ}. The subsequent orbital evolution may (very
roughly) be divided into three stages. In the first and longest stage the orbit
is extremely eccentric, and GWs are emitted in short ``pulses'' during pericenter
passages. These GW pulses slowly remove energy and angular momentum from the
system, and the orbit gradually shrinks and circularizes. After $\sim 10^3-10^8$
years (depending on the two masses and the initial eccentricity) the evolution
enters its second stage, wherein the orbit is sufficiently circular that the
emission can be viewed as continuous.
Finally, as the object reaches the last stable orbit (LSO), the adiabatic
inspiral transits to a direct plunge, and the GW signal cuts off.
Radiation reaction quickly circularizes the orbit over the inspiral; however,
initial eccentricities are large enough that a substantial fraction of captures
will maintain high eccentricity up until the final plunge. (It has been
estimated \cite{BC} that around half the captures will plunge with $e\gtrsim 0.2$.)
While individually-resolvable captures will mostly be detectable during
the last $\sim 1-100$ yrs of the second stage (depending on the CO and MBH
masses), radiation emitted during the first stage will contribute significantly
to the confusion background.

Our first goal will be to estimate the ambient GW energy spectrum arising
from all capture events in the history of the universe. (Later in this paper
we shall be concerned with how much of this GW energy is associated with sources
that are not individually resolvable.) We accomplish this by the following steps:
First, we
devise a model power spectrum for individual captures (with given MBH mass
and LSO eccentricity), based on the approximate, post-Newtonian waveform family
introduced in Ref.\ \cite{BC}. We normalize the overall amplitude of this power
spectrum using a (fully-relativistic) estimate of the total power emitted in GW
throughout the entire inspiral. Next, we introduce an ``average'' power
spectrum, by roughly weighting the individual spectra against (best available
estimates of) the astrophysical distributions of sources' intrinsic parameters,
i.e., the MBH's mass
and spin and the LSO eccentricity and inclination angle. In the last step,
we combine this ``average'' spectrum with astrophysical event rates for captures,
and obtain an expression for the spectral density of the entire capture background.
This translates immediately to an expression for an effective LISA noise spectral
density due to this capture background. For simplicity (and since, in the case of
stellar BHs, the mass distribution is currently poorly modeled),
we shall ``discretize'' CO masses by lumping them into three classes: $10 M_{\odot}$
BHs, $1.4 M_{\odot}$ NSs, and $0.6 M_{\odot}$ WDs.
%We shall also assume identical
%shapes for the GW spectra from these three classes (this may not be strictly the
%case since the actual distribution of initial pericenter distances may differ
%for the various CO species, which however hard to asses).

The total energy output from any given inspiral orbit in Kerr can be estimated
fairly easily even within a full GR context: It is well
approximated by the ``binding energy'' of the CO at the LSO, $m-E$, where
$m$ is the CO's mass, and $E$ is the energy associated with the last
stable geodesic. (The energy radiated during the brief final plunge is negligible.
The energy going down the hole during the inspiral is less than $1\%$ of $m$ in all
cases \cite{Hughes}, and we shall neglect it here as well.)
We find \cite{BCconf} that for astrophysically relevant inspirals with $e_{\rm LSO}
\protect\lesssim 0.35$, the CO emits between $\sim 4\%$ and $\sim 12\%$
of its mass in GWs, depending on the MBH spin and orbital inclination angle $\iota$.
Averaging over inclination angles, assuming orbits are randomly distributed in
$\cos\iota$, we find that captures release $\sim 5$--$7\%$ of $m$ in GWs, depending
on the MBH spin. We shall denote this ``average'' amount of total emitted energy
(expressed as a fraction of $m$) by $\alpha$, and retain it as an unspecified
parameter, with the fiducial value of $0.06$.

To write down a spectrum $f(dE/df)$ for the GW energy emitted over the
inspiral is a more challenging task, as it requires knowledge of the precise
orbital evolution, including a full-GR account of the radiation reaction effect.
The necessary theoretical tools have only recently been developed \cite{BO},
but, at present time, we are still lacking a working code for implementing
this theoretical framework for generic orbits in Kerr.
In the absence of accurate waveforms (and since astronomical capture rates are anyway
sufficiently uncertain---see below), we implement here a much cruder approach to this
problem: We derive the spectrum using the approximate, analytic formalism we
previously developed in \cite{BC}. In this formalism, the overall orbit is
imagined as osculating through a sequence of Keplerian orbits, with rate of change
of energy, eccentricity, periastron direction, etc.\ determined by solving post-Newtonian
(PN) evolution equations. The emitted waveform is determined, at any instant,
by applying the quadrupole formula to an instantaneous Keplerian orbit, using
the expressions derived long ago by Peters and Matthews \cite{PM}.

\begin{figure}[htb]
\centerline{\epsfysize 6cm \epsfbox{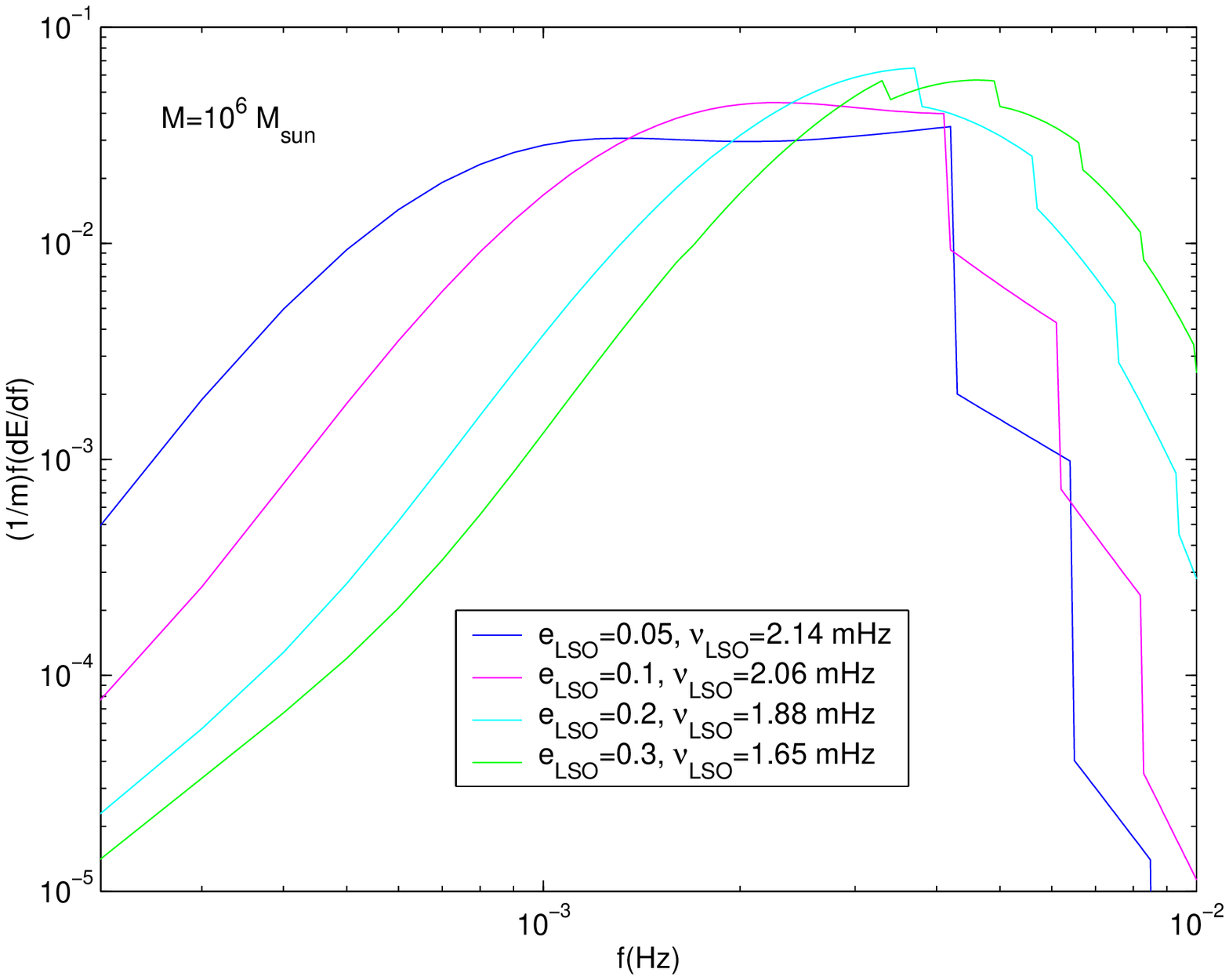}
\epsfysize 6cm \epsfbox{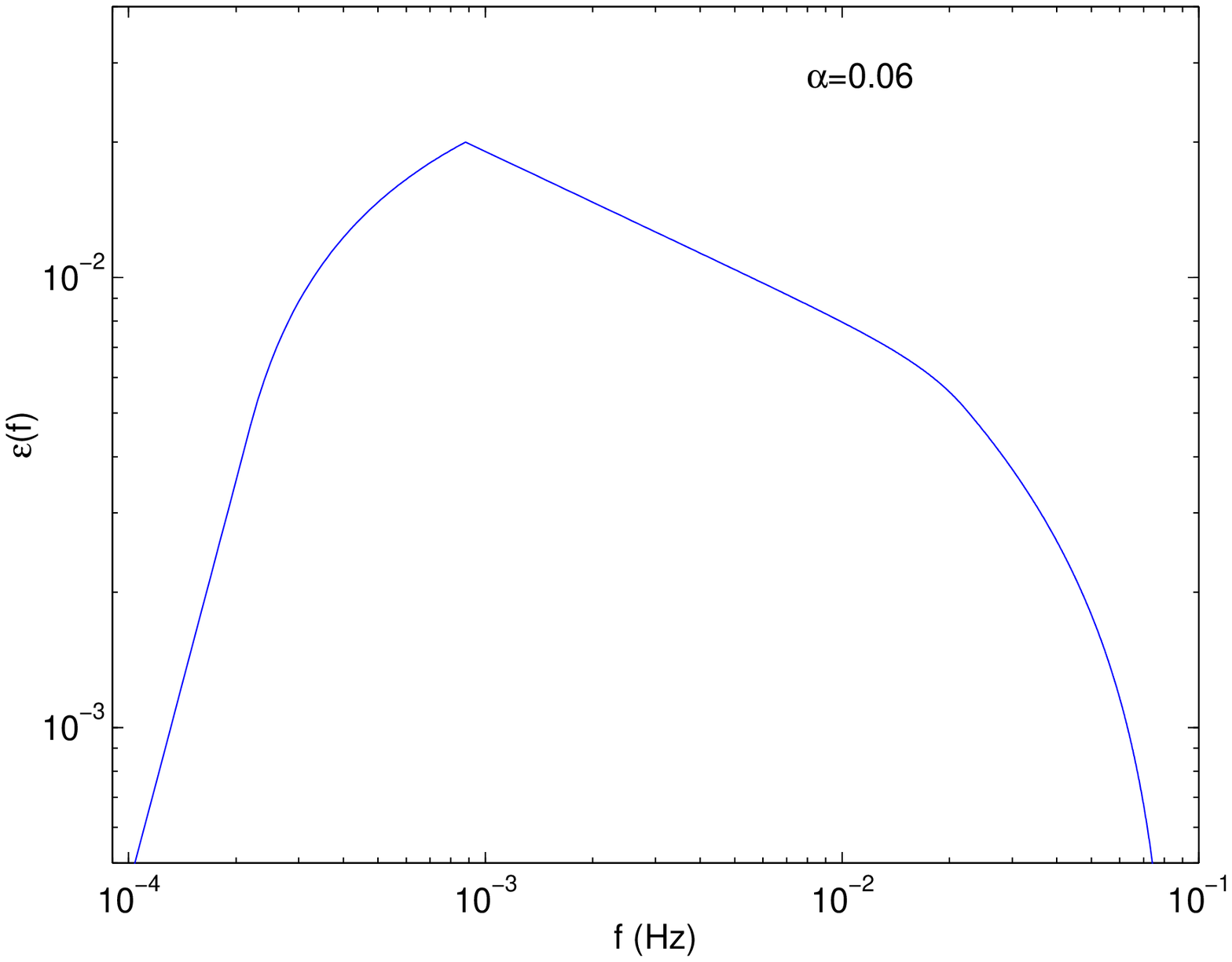}}
\caption{\protect\footnotesize
{\bf Left panel:} Energy spectrum from a single inspiral, for $M=10^6 M_{\odot}$,
$a=0$, and a range of plunge eccentricities $e_{\rm LSO}$. $\nu_{\rm LSO}$ is
the orbital frequency at the LSO. These spectra are based on the approximate,
post-Newtonian waveforms devised in \protect\cite{BC}. Each of the curves represents
the sum of contributions from the first 20 harmonics of the orbital frequency.
``Discontinuities'' appear whenever an $n$-harmonic reaches frequency
$n\times\nu_{\rm LSO}$ and cuts off.
{\bf Right panel:}
Model energy spectrum for a single capture, averaged over MBH mass and
normalized to $\alpha=0.06$.
The apparent sharp drops in the spectrum
at frequencies below $\sim 10^{-3}$ Hz and above $\sim 2\times 10^{-2}$ Hz are
simply an artifact of our restricting attention to MBH masses in the range
$10^{5}$--$10^{7} M_{\odot}$.
}
\label{fig-spectrum}
\end{figure}

More specifically, we apply the following procedure:
For each prescribed value of $e_{\rm LSO}$ we integrate
PN evolution equations [Eqs.~(28) and (30) in \cite{BC}] backwards in time to obtain
the eccentricity $e(t)$ and orbital frequency $\nu(t)$, and, consequently, $e(\nu)$.
We then obtain the power radiated into each of the harmonics of the orbital
frequency (labeled $n=1,2\ldots$) using the leading-order formula \cite{PM}
$
\dot E_n (\nu)= \frac{32}{5} m^2M^{4/3}(2\pi \nu)^{10/3} g_n[e(\nu)],
$
where $g_n(e)$ are certain functions given explicitly in \cite{PM}.
Finally, replacing $\nu\to f_n/n$, we reexpress $\dot E_n(\nu)$ in terms of the
$n$-harmonic GW frequency $f$, and sum up the power from all (in practice,
first 20) harmonics, holding $f$ fixed. Figure \ref{fig-spectrum} (left panel)
shows the single-capture spectrum arising from this procedure, for a
range of $e_{\rm LSO}$ values.
%(The ``discontinuities'' apparent is the spectrum are due to different harmonics
%``cutting off'' at different frequencies, $f_{\rm LSO}=n\,\nu_{\rm LSO}$.)
Note that during the inspiral the source evolves significantly in both
frequency and eccentricity (see, e.g., Figs. 7 and 8 in \cite{BC}), with the GW power
distribution shifting gradually from high harmonics to lower harmonics.
This leads to a spectrum with a steep rise followed by a ``plateau'',
as manifested in Fig.\ \ref{fig-spectrum}.
%For comparison, a decaying, quasi-circular orbit would yield
%a simple power-law spectrum, $f(dE/df)\propto f^{2/3}$.

The fine details of the above spectrum will have little effect on the final
all-capture background spectrum, since averaging over MBH masses (in the
following step) will ``smear out'' these fine details anyway.
The only features that {\em will} be essential are (i) the steep rise at $f\lesssim f_p$,
(where $f_p=2.20$ mHz$/M_6$, $M_6\equiv M/10^6 M_{\odot}$), (ii) the cutoff at $f\sim 4 f_p$,
and (iii) the total energy content. We thus feel justified in approximating the
above shape by a simple ``trapezoidal'' profile, with an overall amplitude normalized
by the known total energy output $\alpha m$ [cf.\ Eq.\ (15) and Fig.\ 4 of \cite{BCconf}].

Next, we consider the ``average'' of our model spectrum over MBH mass $M$, weighted
by the space number density and capture rate for mass $M$. We restrict attention
to MBH masses in the range $0.1\lesssim M_6\lesssim 10$, which
may contribute to the capture background in LISA's most sensitive band, 1--10 mHz.
For these masses, the number density of MBHs, per logarithmic mass interval, has been
estimated as \cite{AR} $dN/d\log M= 2\times 10^{6}\,\gamma\, h_{70}^2\,{\rm Gpc}^{-3}$,
where $h_{70}=H_0/(70$ km\,s$^{-1}$\,Mpc$^{-1}$) and $\gamma\sim 2$ or
$\sim 1$, depending, respectively, on whether or not Sc-Sd host galaxies are
included in the sample \cite{emri}. We retain $\gamma$ as an unknown
factor of order unity. {\em Capture rates} shall be the main source of uncertainty in
our analysis. In \cite{BCconf} we cite a present-time rate (number of captures
per unit proper time per galaxy) of
${\cal R}_0^{A}(M)= \kappa^{A} M_6^{3/8} {\rm yr}^{-1}$,
for captures of CO species `A' ($=$WD, NS, or BH) by a MBH of mass $M=10^6M_{\odot}M_6$.
The (species-dependent) scaling factors $\kappa^A$ are estimated from Freitag's
simulations of the Milky Way \cite{Freitag,FreitagApJ}, but a cautious analysis must
allow for large uncertainties here \cite{emri,Sigurdsson_03}. We shall retain
$\kappa^A$ as unspecified parameters, with probable ranges \cite{emri}
$4\times 10^{-8}\leq \kappa^{\rm WD}\leq 4\times 10^{-6}$ and
$6\times 10^{-8}\leq \kappa^{\rm NS},\kappa^{\rm BH}\leq 6\times 10^{-7}$.

Thus, the (present day) space number density multiplied by the capture rate
scales as $\propto M^{3/8}$, which we now use as the appropriate weight when averaging
our single-capture spectral density over MBH mass (in the range $10^5$ to
$10^7 M_\odot$). This yields a (present day) ``average'' single-capture spectral
density $\epsilon(f)$ as shown in the right panel of Fig.\ \ref{fig-spectrum}
[Eq.\ (22) of
\cite{BCconf}]. Note that the apparent sharp drops in the spectrum
at frequencies below $\sim 10^{-3}$ Hz and above $\sim 2\times 10^{-2}$ Hz are
simply an artifact of our restricting attention to MBH masses in the range
$10^{5}$--$10^{7} M_{\odot}$. In the important intermediate frequency band
the spectral profile is given approximately by
$\epsilon(f)=0.02(\alpha/0.06)(f/1 {\rm mHz})^{-3/8}$.
%------------------------------    FIGURE I  -------------------------------
%----------------------------------------------------------------------------

We now combine the above spectrum with rate estimates to yield the total
energy spectrum of capture background (per species `A'),
$d\rho^A/d\ln f$. As captures are seen to cosmological distances,
we must make here concrete assumptions regarding cosmology and MBH evolution.
[Both MBH mass function and capture rates depend on $z$ (in a way that is
currently poorly modeled); in addition, one must of course account for
cosmological red-shift when integrating the energy density contributions
out to cosmological
distances.] In \cite{BCconf} we argue that, for a range of plausible cosmological
and evolutionary scenarios, the total capture background spectrum can always
be expressed as
$d\rho^A/d\ln f=m^A {\cal R}_0^A T_{\rm eff}\, \epsilon(f)$,
where ${\cal R}_0^A$ is the total preset-day capture rate, and the details
of the specific scenario are encoded only in the value of the effective
integration time
$T_{\rm eff}$. Thus, a cosmological/evolutionary scenario with a certain
$T_{\rm eff}$ is equivalent, as far as $d\rho^A/d\ln f$ is
concerned, to a flat-universe/no-evolution model, in which all captures
had been ``turned on'' $T_{\rm eff}$ years ago. We further show \cite{BCconf}
that plausible cosmological/evolutionary scenarios all agree with the
flat-universe/no-evolution model to within a mere $15\%$ if one makes the
judicious choice $T_{\rm eff}=7\times 10^9$ years. We retain $T_{\rm eff}$
as an unspecified parameter, with a fiducial value of $7\times 10^9$ years.

\begin{figure}[htb]
\centerline{\epsfysize 6cm \epsfbox{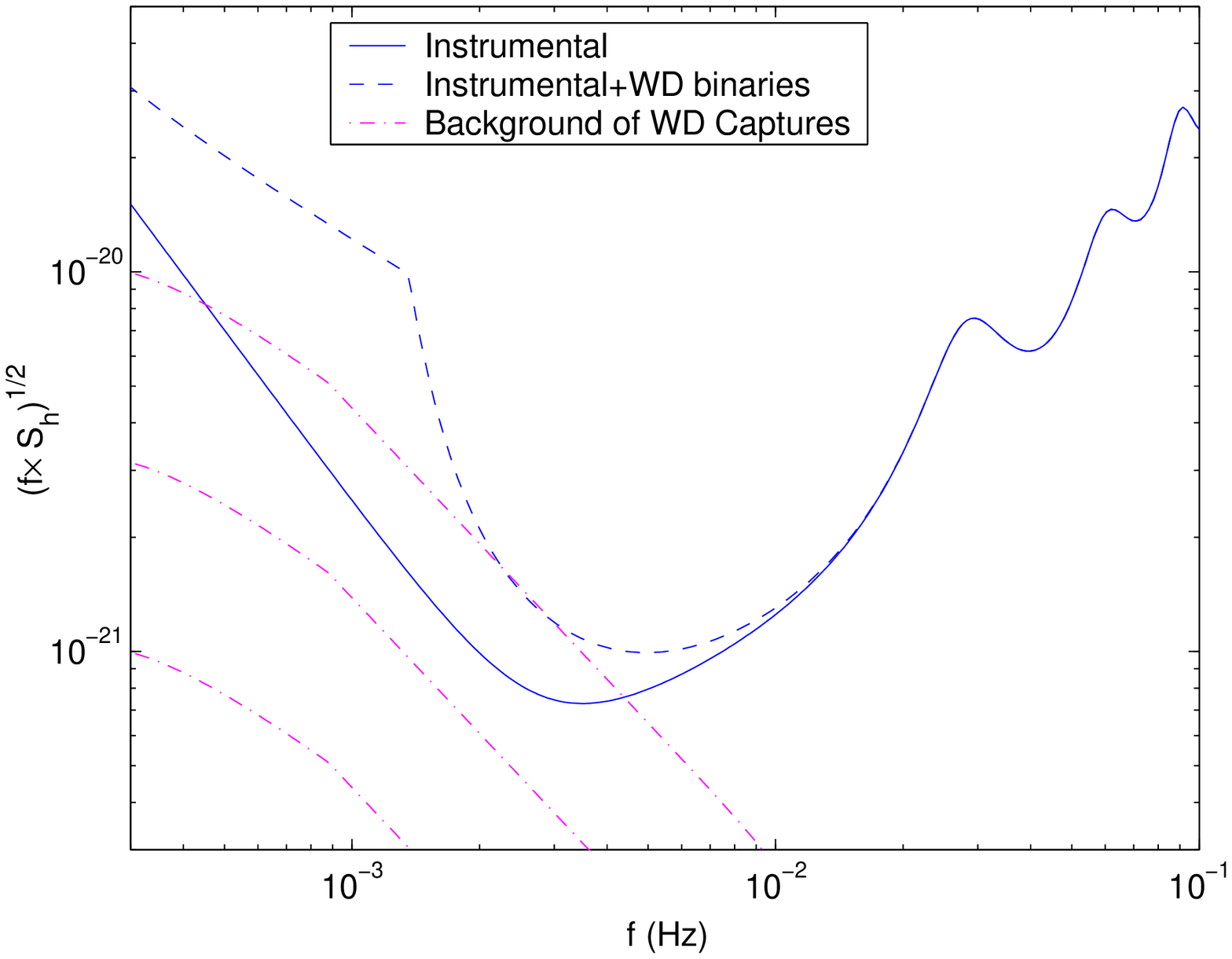}
\epsfysize 6cm \epsfbox{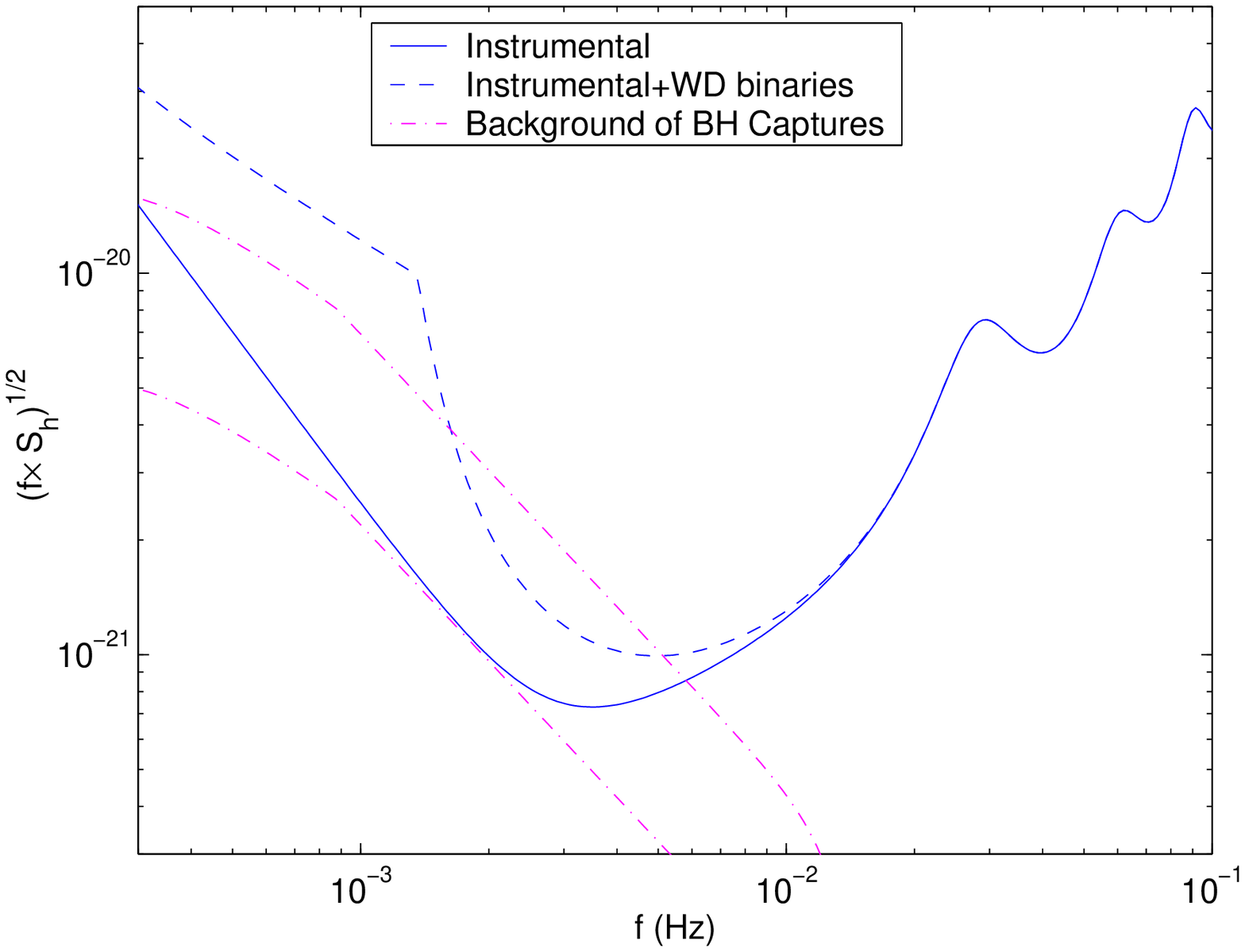}}
\caption{\protect\footnotesize
{\bf Left panel:}
Comparison of the background ${\cal S}_h^{\rm WDcapt}$ from all WD captures
(dash-dot line) with LISA's instrumental noise (solid line) and with
LISA's instrumental plus WD-binaries confusion noise (dashed line).
We show three cases, corresponding to WD capture rates of
$\kappa^{WD} = 4\times 10^{-6}, 4\times 10^{-7}$, and $4\times 10^{-8}$,
respectively (see text).
{\bf Right panel:}
Same as left panel, for the background ${\cal S}_h^{\rm BHcapt}$ from all BH
captures. We show two cases, corresponding to
BH capture rates of $\kappa^{\rm BH} = 6\times 10^{-7}$ and
$6\times 10^{-8}$.
}
\label{fig:fullnoise}
\end{figure}

There will likely be tens of thousands of CO capture sources in the LISA band
at any instant, most of which---as we shall see---unresolvable. It is then
justified to think of the ambient GW energy from captures as constituting an
isotropic background. As such, the capture background represents (for the
purpose of analyzing {\it other} sources) a noise source with spectral
density~\cite{stochUL}
${\cal S}^{\rm capt}_h(f) = \frac{4}{\pi} f^{-3}d\rho/d\ln f$.
(For important comments regarding notational conventions for LISA noise model,
see Sec.\ IV-B of \cite{BCconf}.)
Substituting for $d\rho/d\ln f$ we finally obtain the desired
noise spectral density from the capture background---see Eq.\ (33) of \cite{BCconf}.
In the crucial band $1\, {\rm mHz}\lesssim f\lesssim 20\, {\rm mHz}$, we find
%~~~~~~~~~~~~~~~~~~~~~~~~~~~~~~~~~~~~~~~~~~~~~~~~~~~~~~~~~~~~~~~~
\begin{eqnarray} \label{S}
{\cal S}^{\rm Acapt}_h(f)\approx
8\cdot 10^{-40} \ {\rm Hz}^{-1}
\left(\frac{m^A}{M_{\odot}}\right)
\left(\frac{\alpha}{0.06}\right)\left(\frac{T_{\rm eff}}{7\cdot 10^9 {\rm yr}}
\right)
\left(\frac{\kappa^{A}}{10^{-7}}\right)\,\gamma\, h_{70}^2
\left(\frac{f}{1\; {\rm mHz}}\right)^{-27/8}.
\end{eqnarray}
%~~~~~~~~~~~~~~~~~~~~~~~~~~~~~~~~~~~~~~~~~~~~~~~~~~~~~~~~~~~~~~~~
In Fig.\ \ref{fig:fullnoise} we depict this all-captures background,
for the sake of comparing with LISA's instrumental noise. We show
${\cal S}^{\rm Acapt}_h$ for
different values of $\kappa^A$, in the ranges specified above. (For lack
of space, we present here results for WD and BH captures only; the
corresponding plots for NS captures---whose contribution to the confusion
background, anyhow, turns out to be negligible---can be found in \cite{BCconf}.)
%------------------------------    FIGURE VI  -------------------------------
%--------------------------------------------------------------------------

Now, ${\cal S}^{\rm capt}_h(f)$ is the spectrum of the background from
{\it all} captures, and thus represents an upper limit on the effect of
confusion from captures. Since, near the floor of the LISA noise curve,
this upper limit may surpass LISA's instrumental noise (assuming the high
end of the estimated rates), it is important to next consider what fraction of
${\cal S}^{\rm capt}_h(f)$ actually constitutes an unresolvable confusion
background, or, equivalently, what portion is resolvable and hence subtractable.

The issue of source subtraction is generally a challenging one, and we
discuss it in some detail in \cite{BCconf}. When confusion noise is the
dominant noise source---as may well be the case in reality for capture
confusion---one faces a ``chicken-and egg'' kind of problem: One
cannot determine which sources are detectable without first knowing
the confusion noise level; yet, to calculate the confusion noise level,
one must know which sources are detectable and hence subtractable.
This is a manifestation of the fact that when the event rate is high enough
that confusion noise starts to dominate over other noise sources,
the level of this confusion noise depends non-linearly on the event rate
\cite{BCconf}. This problem does not show up for classes of sources
intrinsically weak enough that even at a very low event rate only a tiny
portion of the associated confusion energy is resolvable. Such turns out
to be the case with WD and NS captures. The situation with BH captures
is more involved, as in this case it is likely that a substantial portion
of the confusion energy {\em will} be resolvable. As a consequence,
the estimate we give below for the unsubtractable part of
${\cal S}^{\rm capt}_h(f)$ will be very crude in the BH case.

To estimate the subtractable portion of ${\cal S}^{\rm Acapt}_h(f)$
(for each species A), we have referred again to our approximate waveform
model. Based on that model, we estimated (for a range of capture parameters)
(i) how much of the capture GW energy is radiated as a function of time along
the capture history, and (ii) how the SNR output (as seen by LISA, for a source
at a given distance with, say, 3-yr-long matched filters) varies over the
capture history. Cf.\ Figs.\ 11 and 13 of \cite{BCconf}.
Assuming a realistic detection SNR threshold of 30 \cite{emri}, the relation
(ii) translates to an estimate of how {\em detection
distance} depends upon time along a given inspiral. We then crudely average
over capture parameters to get an ``average'' description of how the
contribution to the capture energy density, and the detection distance,
both vary over time. This information, along with the fact that an estimated
$80\%$ of the capture background comes from $z<1$ (see the discussion in Sec. IV-D
of \cite{BCconf}), suffices for our rough estimation:
For all sources that (during LISA's observation time) have a certain time
to go until their final plunge, we know what the average detection distance is,
and so (assuming sources are isotropically distributed) what portion of the
background energy from these sources is subtractable. Weighting this by the
relative contribution from these sources to the overall capture energy, and
integrating over the entire inspiral, gives the overall subtractable fraction of energy.
(This is merely the essential idea of our estimate; see \cite{BCconf} for more details.)
Denoting the {\em unsubtractable} confusion noise for CO spices A by
$S_h^{\rm Acapt}(f)$ (note: ``calligraphic'' typeface for the total background vs.\
``upright'' typeface for the unsubtractable confusion portion), we estimate
\begin{equation}\label{Sconf}
S_h^{\rm WDcapt} \approx 0.97\times {\cal S}_h^{\rm WDcapt},
\quad\quad
S_h^{\rm NScapt} \approx 0.94\times {\cal S}_h^{\rm NScapt},
\quad\quad
S_h^{\rm BHcapt} = (0.3 - 1) \times {\cal S}_h^{\rm BHcapt},
\end{equation}
where in the BH case we are forced to extend the range upward to include
the possibility that $S_h^{\rm BHcapt}$ significantly raises the
total effective noise level (our analysis is not sufficient to narrow
the range of possibilities in this case).
The estimates quoted here change very little if one assumes SNR thresholds
of 15 or 60 (instead of 30) \cite{BCconf}. In any case, it seems justified
to approximate $S_h^{\rm WDcapt} \approx {\cal S}_h^{\rm WDcapt}$
and $S_h^{\rm NScapt} \approx {\cal S}_h^{\rm NScapt}$, while in
the BH case we leave it to future work to improve our crude estimate.
We finally note that the above estimate ignores the obvious $f$ dependence
of the ratio ${\cal S}_h^{\rm NScapt}(f)/S_h^{\rm NScapt}(f)$, and only
concerns the overall capture energy content. This is a valid approximation
if most of the capture energy from the relevant confusion sources is emitted
at frequencies near LISA's floor sensitivity band. That this indeed is the
case, is discussed in \cite{BCconf} (see Sec.\ V-C and Fig.\ 12 therein).

\begin{figure}[htb]
\centerline{\epsfysize 6cm \epsfbox{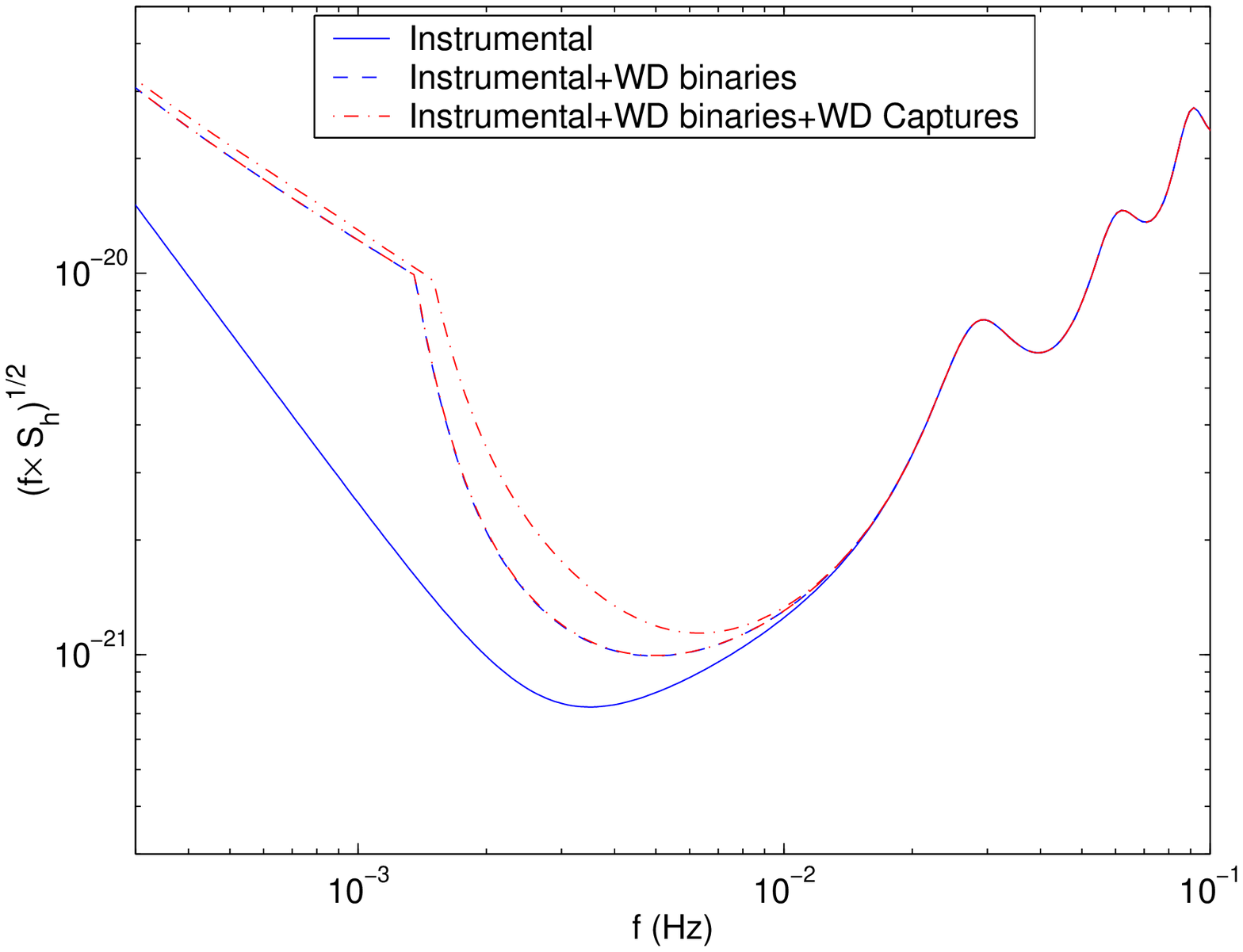}
\epsfysize 6cm \epsfbox{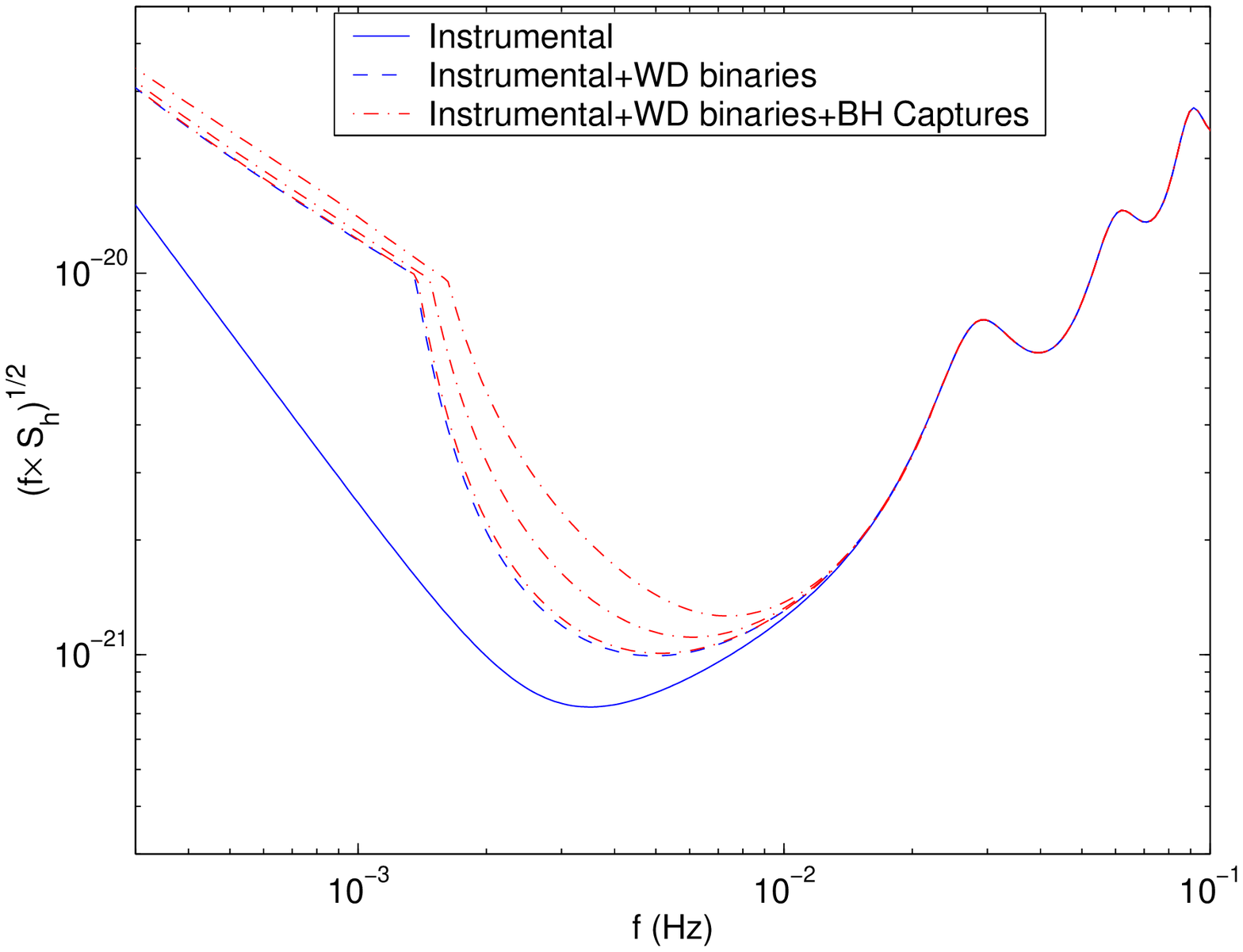}}
\caption{\protect\footnotesize
{\bf Left panel:}
Total LISA noise (dash-dot line), including instrumental noise,
confusion from WD binaries, and confusion from WD captures.
The two total-noise curves correspond to the higher and lower ends of
the estimated event rate for WD captures.
We have assumed here that none of the WD capture background is subtractable.
{\bf Right panel:}
Same, with confusion noise from BH captures.
The three dash-dot lines show the total noise curve under different
assumption as to the event rate $\kappa^{\rm BH}$ and the fraction of
subtractable noise.
The lower total-noise curve refers to the {\em lower}
end of event rate estimates ($\kappa^{\rm BH} = 6 \times 10^{-8}$),
with $S_h^{\rm BHcapt}$ assumed to be $0.3\times {\cal S}_h^{\rm BHcapt}$.
The middle total-noise
curve corresponds to the {\em upper} end of rate estimates
($\kappa^{BH} = 6 \times 10^{-7}$), again with
$S_h^{\rm BHcapt} = 0.3\times {\cal S}_h^{\rm BHcapt}$.
The upper curve assumes the upper end of rate estimates, but with
$S_h^{\rm BHcapt}= {\cal S}_h^{\rm BHcapt}$.
Thus the upper curve represents an upper
limit on the effect of confusion noise from BH captures.
}
\label{fig:Seff}
\end{figure}

Lastly, we describe the effect of the (unsubtractable) capture confusion noise
on the LISA noise curve. Capture noise does not simply add in quadrature to
the other noise sources. Roughly speaking, this is because in the crucial 2--5 mHz
band LISA's effective noise level is dominated by ``imperfectly subtracted''
galactic WD binaries (GWDB), which reduce the bandwidth available for other
types of sources, and hence effectively magnify the capture confusion noise.
This effect taken into account, the total effective spectral noise density is
approximated by \cite{Hughes02,BCconf}
\begin{eqnarray}\label{Stot}
S^{\rm eff}_h(f) = {\rm min}
\left\{  \left(S_h^{\rm inst} + {\cal S}_h^{\rm EGWDB} + S_h^{\rm capt}\right)
\exp(\kappa T^{-1} dN/df),\;
S_h^{\rm inst} + {\cal S}_h^{\rm GWDB} + {\cal S}_h^{\rm EGWDB} +
S_h^{\rm capt}\right\},
\end{eqnarray}
where
$S_h^{\rm inst}$ is LISA's instrumental noise \cite{Larson}, and
${\cal S}^{\rm GWDB}_h= 1.4\cdot10^{-44}f^{-7/3}/{\rm Hz}$
and ${\cal S}_h^{\rm EGWDB}=2.8 \cdot 10^{-46} f^{-7/3}/{\rm Hz}$
(where $f$ is in Hz) are estimated spectral densities for galactic and
extra-galactic WD binaries, respectively \cite{FarmerPhinney,Nelemans_2001c}.
In the exponent, $dN/df= 2\cdot10^{-3}f^{-11/3}/{\rm Hz}$ ($f$ in Hz)
estimates the number density of GWDBs per unit GW frequency \cite{Hughes02},
$T$ is the LISA mission lifetime, and $\kappa$ is the average number of frequency
bins that are ``lost'' (for the purpose of analyzing other sources) when each GWDB
is fitted out. We take $\kappa T^{-1} = 1.5/{\rm yr}$ \cite{BCconf}.
In Fig.\ \ref{fig:Seff} we depict $S^{\rm eff}_h(f)$,
with the contributions from the WD and BH captures considered separately
(again, we refer the reader to \cite{BCconf} for the NS case).
For WDs we have included {\em all} capture noise as confusion noise,
as suggested above.
For BHs we show two cases where the nonsubtractable portion is assumed
to be 30\% of the background, plus one case (at the upper end for capture rates)
where the nonsubtractable fraction is assumed to be 100\% of the background.
Note that the astrophysical event rate remains the main source of uncertainty
in our analysis, and clearly overwhelms the uncertainty introduced by our
crude estimate of the ratio $S_h^{\rm capt}/{\cal S}_h^{\rm capt}$.

%------------------------------    FIGURE XIV  -------------------------------
%--------------------------------------------------------------------------

We conclude that the effect of capture confusion noise on the total LISA
noise curve is rather modest: Even for the highest capture rates
we consider, the total LISA noise level $[f S^{\rm eff}_h(f)]^{1/2}$ is
raised by a factor $\lesssim 2$ in the crucial frequency range 1--5 mHz.
While this slightly elevated noise level would somewhat decrease LISA's
sensitivity to {\it other} classes of sources, we note that, overall, this
is a pleasant situation, as it also implies that capture detection
rates are at nearly their maximum possible levels.
Roughly, the reason is that detection rate for captures, $\cal D$, considered as a
function of the astrophysical event rate, $\cal R$ (for a given detector design and
a given level of confusion noise from GWDBs), must peak at about the rate
where capture confusion starts to dominate over instrumental noise, since
for a much lower rate $\cal D$ clearly grows with $\cal R$, whereas for a rate
much higher (at which capture confusion dictates the detection distance) $\cal D$
should clearly drop with $\cal R$. Again, we refer the reader to \cite{BCconf}
for a more detailed discussion.

{\bf Acknowledgements:}
We thank the members of LIST's Working Group 1, and especially Sterl Phinney,
from whom we first learned of the problem.
We also thank Marc Freitag for helpful discussions of his capture simulations.
C.C.'s work was partly supported by NASA Grant NAG5-12834.
L.B.'s work was supported by NSF Grant NSF-PHY-0140326 (`Kudu'),
and by a grant from NASA-URC-Brownsville (`Center for Gravitational
Wave Astronomy'). L.B. acknowledges the hospitality of the Albert Einstein
Institute, where part of this work was carried out.

\end{document}